\documentclass[prd,showpacs,preprintnumbers]{revtex4} % Physical Review D

\usepackage{graphicx}% Include figure files
\usepackage{dcolumn}% Align table columns on decimal point
\usepackage{bm}% bold math
\usepackage{latexsym}
\usepackage{amsfonts}
\usepackage{amssymb}
\usepackage{amsmath}
\begin{document}
\preprint{OU-HET-1050, KOBE-COSMO-20-06}
\title{Polarized initial states of primordial gravitational waves}
\author{Sugumi Kanno}
\email{sugumi@het.phys.sci.osaka-u.ac.jp}
\author{Jiro Soda}
\email{jiro@phys.sci.kobe-u.ac.jp}
\affiliation{Department of Physics, Osaka University, Toyonaka 560-0043, Japan$^{*}$\\
Department of Physics, Kobe University, Kobe 657-8501, Japan$^{\dag}$}
\date{\today}

%===============================================================%
%************************* ABSTRACT ****************************%
%===============================================================%
\begin{abstract}
Polarizations of primordial gravitational waves can be relevant when considering inflationary universe in modified gravity or when matter fields survive during inflation. Such polarizations have been discussed in the Bunch-Davies vacuum. Instead of taking into account dynamical generation of polarizations of gravitational waves, in this paper, we consider polarized initial states constructed from $SU(2)$ coherent states. We then evaluate the power spectrums of the primordial gravitational waves in the states.\end{abstract}

\maketitle

\section{Introduction}

The inflationary universe has succeeded in resolving conceptual puzzles such as the flatness problem and explaining   observational data  of
the cosmic microwave background radiation and the large scale structure of the universe. The most surprising prediction of the inflationary universe
is that the origin of cosmic structure stems from quantum fluctuations. However, any compelling observational evidence of the quantum nature of the origin has not yet been found. Hence, to find the quantum nature of the origin of primordial fluctuations will be a decisive evidence of the inflationary universe. One of the ideas of hunting the quantum nature of the origin would be to make use of primordial gravitational waves generated directly from the quantum fluctuations of spacetime~\cite{Kanno:2018cuk, Kanno:2019gqw}. 
Since the quantum nature of the origin is also encoded in the initial quantum state,
it is important to understand the initial quantum state of primordial fluctuations. 

In the conventional inflationary universe, the de Sitter invariant Bunch-Davies vacuum is chosen as the initial quantum state for curvature perturbations. 
Interestingly, the Bunch-Davies vacuum can be regarded as a squeezed state
from observers in radiation or matter dominated era. 
Mathematically, however, the de Sitter invariance does not select the initial state uniquely. In fact, there are infinite number of invariant states, $\alpha$-vacua, which are excited states from the point of the Bunch-Davies vacuum.
Moreover, coherent states~\cite{Glauber:1963fi,Koh:2004ez} can be induced in the presence of matter fields during inflation~\cite{Kanno:2018cuk, Kanno:2019gqw}. 

Similarly for the primordial gravitational waves, we usually choose the Bunch-Davies vacuum although we can choose $\alpha$-vacua or coherent states alternatively as the initial quantum state. In the case of primordial gravitational waves, however, they have two degrees of freedom of polarizations unlike the curvature perturbations with one degree of freedom. Since the primordial gravitational waves have not been detected yet, the initial quantum state could be different from that of curvature fluctuations. In order to incorporate the degrees of freedom of polarizations in the initial quantum state, we are able to consider a polarized coherent state as an alternative initial quantum state. We will see that there exists the degree of degeneracy in the total number of gravitons due to the degrees of freedom of polarizations and this can be understood as hidden $SU(2)$ degeneracy. Thus it would be viable to consider 
Sch\"odinger type coherent states constructed from $SU(2)$ coherent states~\cite{Moran, Hussin}. Since the state incorporates the degrees of polarizations, we call the states polarized initial states in this paper. We then calculate the power spectrums of primordial gravitational waves in the polarized initial state for observations.

The organization of this paper is as follows.
In section II, we review how to quantize gravitational waves in the conventional inflationary universe.
In section III, we introduce $SU(2)$ coherent states and Schr\"{o}dinger type coherent states.
In section IV, we study the Shr\"odinger type coherent state and find that
the state describes polarized states of primordial gravitational waves.
We then calculate the power spectrum of primordial gravitational waves for observations. The final section is devoted to the discussion.

\section{Conventional initial states of fluctuations in inflationary cosmology}
\label{section2.1}

In inflationary cosmology, the Bunch-Davies vacuum is usually assumed as the simplest initial state of quantum fluctuations of the universe. This is because spacetime looks flat at short distances and quantum fluctuations are expected to start in a minimum energy state. 

The metric of tensor perturbations is expressed as
\begin{eqnarray}
ds^2=a^2(\eta)\left[-d\eta^2+(\delta_{ij}+h_{ij}) dx^idx^j\,\right] \,,
\label{metric}
\end{eqnarray}
where $\eta$ is the conformal time, $x^i$ are spatial coordinates, $\delta_{ij}$ and $h_{ij}$ are the Kronecker delta and the tensor perturbations which satisfy $h_{ij}{}^{,j}=h^i{}_i=0$. The indices $(i,j)$ run from $1$ to $3$.  The scale factor $a(\eta)$ in inflationary epoch approximated by de Sitter space is given by
\begin{eqnarray}
a(\eta)=-\frac{1}{H \eta }\,,\qquad\eta<0\,.
\end{eqnarray}
Substituting the metric Eq.~(\ref{metric}) into the Einstein-Hilbert action, we obtain the quadratic action
\begin{eqnarray}
 \frac{M_{\rm pl}^2}{2}\int{\rm d}^4x \sqrt{-g}\,R
= \frac{M_{\rm pl}^2}{8}\int{\rm d}^4x\,a^2
\left[\,h^{ij\,\prime}h^\prime_{ij}-h^{ij,k}\,h_{ij,k}
\,\right]  \, ,
\end{eqnarray}
where $M_{\rm pl}^2=1/(8\pi G)$ and a prime denotes the derivative with respect to the conformal time.
We can expand the metric field $h_{ij}(\eta, x^i)$ in terms of the Fourier modes 
\begin{eqnarray}
a(\eta) h_{ij}(\eta, x^i) = {\frac{\sqrt{2}}{M_{\rm pl}}}\frac{1}{\sqrt{V}}\sum_{\bm k}\sum_{A} \ h_A^{\bm{k}}(\eta)\,e^{i {\bm k} \cdot {\bm x}} \ p_{ij}^A(\bm k)  \,,
\label{fourier}
\end{eqnarray}
where we introduced the polarization tensor $p^A_{ij}({\bm k})$ normalized as $p^{*A}_{ij} p^B_{ij} =2 \delta^{AB}$. 
Here, the index $A$ denotes the linear polarization modes $A=\oplus,\otimes$. 
Note that  we consider finite volume $V=L_{x }L_{y}L_{z}$ and 
 discretize the ${\bm k}$-mode with a width ${\bm k} = \left(2\pi  n_x/L_x\,,2\pi  n_y/L_y\,,2\pi  n_z/L_z\right)$ where ${\bm n}$ 
are integers.  

In quantum field theory, the metric field in the right hand side, $h_A^{\bm k}(\eta)$, is promoted to the operator.
The  operator $h_A^{\bm k}$ satisfies
\begin{eqnarray}
h_{A}^{\bm k\prime\prime }+\left(k^2-\frac{a''}{a}\right)  h_A^{\bm k}=0\,.
\label{mf}
\end{eqnarray}
In the inflationary era, the operator $h^A_{\bm k}(\eta)$ is then expanded in terms of creation and annihilation operators
\begin{eqnarray}
h_A^{\bm k}(\eta)=b_A^{\bm k}\,v_k(\eta)+b^{-{\bm k}\dag}_{A }\,v^*_k (\eta)\,,
\label{operatorsI}
\end{eqnarray}
where 
$
\left[b_A^{\bm k} , b^{{\bm p}\dag}_{B} \right]
= \delta_{AB} \delta^{\bm k,\bm p}
$ is satisfied, $k$ is the magnitude of the wave number ${\bm k}$, and $*$ denotes complex conjugation.
Eq.~(\ref{mf}) gives the positive frequency modes $v_k$ in the form
\begin{eqnarray}
v_{k} (\eta) \equiv \frac{1}{\sqrt{2k}}\left(1-\frac{i}{k  \eta}\right)e^{-ik  \eta }
\,.
\label{positivefreq}
\end{eqnarray}
The Bunch-Ddavies vacuum is defined as
\begin{eqnarray}
b^{\bm k}_A\,|0 \rangle =0   \ .
\label{vacua}
\end{eqnarray}

In the Bunch-Davies vacuum, the equations of motion for different polarization modes ($\oplus,\otimes$) are decoupled and  exactly the same in the absence of sources and we don't need to discuss both modes. However, if the symmetry of de Sitter invariance breaks down due to other matter fields, we might need to distinguish the mode functions $v_k^{\rm A}$ between  different polarization modes.

\section{$SU(2)$ coherent states}

The polarizations of gravitational waves are expressed in terms of linearly polarized waves with a plus polarization $h_\oplus$ and a cross polarization $h_\otimes$. They behave independently when no interaction between the gravitational waves and matter fields exists as seen in the previous section. In the presence of the interaction, however, those two polarizations can be mixed. In this section, we consider Schr\"odinger type coherent states constructed from $SU(2)$ coherent states~\cite{Moran, Hussin}, which are assumed to arise from a consequence of the dynamics before inflation. The $SU(2)$ coherent states are not de Sitter invariant. Thus, we assume some excitations in the beggining of the universe. Since the primordial gravitational waves have not been detected yet, the initial quantum state could be different from that of curvature fluctuations.
The index ${\bm k}$ is omitted for simplicity unless there may be any confusion below.

We can define the Fock space for the gravitational waves as 
\begin{eqnarray}
&& b_\oplus |n,m\rangle  = \sqrt{n}\,|n-1, m\rangle \ ,  \qquad
b_\oplus^{\dag} |n,m\rangle = \sqrt{n+1}\,|n+1, m\rangle  \ , 
\label{plus}\\
&& b_\otimes |n,m\rangle  = \sqrt{m}\,|n, m-1\rangle \ ,\qquad
b_\otimes^{\dag} |n,m\rangle  = \sqrt{m+1}\,|n, m+1\rangle 
\label{cross}\ .
\end{eqnarray}
Here, $|n,m\rangle$ represents $n$ gravitons with a polarization $\oplus$
and $m$ gravitons with  a polarization $\otimes$.
The total number of gravitons are given by $\nu = n+m, \ (\nu=0,1,2 \cdots)$ and we see there are degenerate states for a given $\nu$.
We also define creation and annihilation operators that create and annihilate a plus and a cross polarizations with the ratio $\alpha$ to $\beta$ such as
\begin{eqnarray}
&& A_{\alpha ,\beta}^{\dag}  = \alpha\,b_\oplus^\dag  + \beta\,b_\otimes^\dag \ , 
\label{creation} \\
&& A_{\alpha ,\beta}  = \alpha^*b_\oplus + \beta^*b_\otimes  \ ,
\label{annihilation}
\end{eqnarray}
where $\alpha ,\beta$ are complex numbers with normalization 
$|\alpha|^2+|\beta|^2 =1$ so that they satisfy
\begin{eqnarray}
[\,A_{\alpha ,\beta}\,,\,A_{\alpha ,\beta}^{\dag}\,] = 1 \ .
\label{commutation}
\end{eqnarray}
The eigenstate of graviton number $\nu$ that consists of any polarizations is given by
\begin{eqnarray}
   |\nu ,\alpha ,\beta\rangle \equiv \frac{1}{\sqrt{\nu!}} 
            \left( A_{\alpha ,\beta}^{\dag}\right)^\nu |0\rangle
               = \sum_{n=0}^\nu \alpha^n \beta^{\nu -n} \sqrt{{}_\nu C_n} 
   |n,\nu -n \rangle
   \label{eigenstate}
\end{eqnarray}
where we used Eqs.~(\ref{plus}) (\ref{cross}) and (\ref{creation}) and defined the ground state with non-degenerate state $|0,\alpha,\beta\rangle\equiv|0\rangle$. The states $|\nu , \alpha ,\beta\rangle$ are precisely the $SU(2)$ coherent states in the Schwinger boson representation. We see that there exists a degeneracy of $\nu+1$ in the total number of gravitons $\nu$. This makes sense from the fact that
the degeneracy present in the spectrum is an $SU(2)$ degeneracy and
the states are averaged out the degenerate contributions to a given $\nu$.
The states satisfy the relation
\begin{eqnarray}
\langle\mu ,\gamma , \delta\,|\,\nu ,\alpha ,\beta\rangle 
= \left(\gamma^* \alpha + \delta^* \beta \right)^\nu \delta_{\mu ,\nu}    \ .
\end{eqnarray}
It implies
\begin{eqnarray}
\langle\mu ,\alpha ,\beta\,|\,\nu ,\alpha ,\beta\rangle 
= \delta_{\mu ,\nu}    \ .
\end{eqnarray}
due to the normalization $|\alpha|^2+|\beta|^2 =1$.
The operator $A_{\alpha ,\beta}$ ($A_{\alpha ,\beta}^\dag$) decreases (increases) the total number of gravitons by one such as
\begin{eqnarray}
A_{\alpha ,\beta} |\,\nu ,\alpha ,\beta\rangle =\sqrt{\nu}\,|\,\nu -1 ,\alpha ,\beta\rangle  \ , \quad
A_{\alpha ,\beta}^{\dag}\,|\,\nu ,\alpha ,\beta \rangle =\sqrt{\nu +1}\,|\,\nu +1 ,\alpha ,\beta\rangle  \ ,
\end{eqnarray}
where we used Eq.~(\ref{eigenstate}).

Now, we define a Schr\"{o}dinger type coherent state by an eigen equation
\begin{eqnarray}
  A_{\alpha ,\beta} |\Psi , \alpha ,\beta\rangle 
  = \Psi \,|\Psi , \alpha ,\beta\rangle\ ,
\label{schrodinger}
\end{eqnarray}
where $\Psi$ is an eigenvalue. 
Since $A_{\alpha,\beta}$ is not hermitian, $\Psi$ is in general, a complex number $\Psi=|\Psi|e^{i\theta}$ with a phase $\theta$. If we expand $|\,\Psi , \alpha ,\beta\rangle$ in terms of $ |\nu ,\alpha ,\beta\rangle$, we find that the coherent state $|\,\Psi , \alpha ,\beta\rangle$ can be created from the vacuum by the displacement operator $D(\Psi)$ as follows,
\begin{eqnarray}
   |\Psi , \alpha ,\beta\rangle
  &=& e^{-\frac{|\Psi|^2}{2}} \sum_{\nu=0}^\infty \frac{\Psi^\nu}{\sqrt{\nu!}}\,
  |\nu ,\alpha ,\beta\rangle
  = e^{-\frac{|\Psi|^2}{2} +\Psi A_{\alpha ,\beta}^{\dag}} 
  |0 \rangle  \nonumber\\
  &=&  \exp\left[\Psi A_{\alpha ,\beta}^{\dag} -  \Psi^* A_{\alpha ,\beta}\right]
  |0 \rangle  \ \equiv  D(\Psi)|0  \rangle
  \ .
 \label{displacement}
\end{eqnarray}
where we used a relation $\exp{(A)}\exp{(B)}=\exp(A+B+1/2[A,B])$ and Eq.~(\ref{commutation}) in the third equality.
Note that the displacement operator has the following relation
\begin{eqnarray}
    D(\Psi) 
= \exp\left[\alpha\,\Psi\,b_{\oplus}^{\dag} -  \alpha^* \Psi^* b_{\oplus}
  + \beta\,\Psi\,b_{\otimes}^{\dag} -  \beta^*\Psi^* b_{\otimes}\right]
    \equiv D_\oplus (\alpha \Psi ) D_\otimes (\beta \Psi)
  \ .
\end{eqnarray}
This tells us that the eigenvalues for $b_{\oplus}$ and  $b_{\otimes}$ are multiplied respectively by $\alpha$ and $\beta$ as
\begin{eqnarray}
  b_{\oplus} |\Psi , \alpha ,\beta\rangle 
  = \alpha \Psi  |\Psi , \alpha ,\beta\rangle \ , \quad
   b_{\otimes} |\Psi , \alpha ,\beta\rangle 
  = \beta \Psi\,|\Psi , \alpha ,\beta\rangle
  \label{key} \ .
\end{eqnarray}
Notice that the statistical distribution of gravitons in the coherent state is found to be the Poisson distribution 
\begin{eqnarray}
|\,\langle\nu ,\alpha ,\beta\,|\,\Psi ,\alpha ,\beta\rangle\,|^2   
= e^{-|\Psi|^2}  \frac{\Psi^{2\nu}}{\nu!}
   \ .
\end{eqnarray}
where we used Eq.~(\ref{displacement}). Here, we note that there are four degrees of freedom in the Schr\"{o}dinger type coherent state $|\Psi , \alpha ,\beta\rangle$. This is because we have three complex parameters $\alpha , \beta , \Psi$ which are constrained by $|\alpha|^2 +|\beta|^2 =1$. And the phase of $\Psi$ can be absorbed into $\alpha$ and $\beta$. The remaining four degrees of freedom
represent the intensity and three degrees of polarizations respectively as we will see in Eqs.~(\ref{N})$\sim$(\ref{V}).

\section{Polarized initial states of primordial gravitational waves}

In this section, we discuss observations of primordial gravitational waves in the Schr\"odinger type coherent state $|\Psi , \alpha ,\beta\rangle$. Below, we will call $|\Psi , \alpha ,\beta\rangle$ polarized initial state since the state incorporates polarizations as the initial state. To the best of our knowledge, there is no explicit study relating observations to the polarized initial state.

In order to discuss the polarization of primordial gravitational waves, 
let us first define the total number operator $N$ which is proportional to the intensity as
\begin{eqnarray}
N =  b_\oplus^\dag  b_\oplus +  b_\otimes^\dag  b_\otimes \ ,
\label{N}
\end{eqnarray}
and the Stokes operators in the case of gravitational waves~\cite{Kato:2015bye}
\begin{eqnarray}
Q &=& b_\oplus^\dag  b_\oplus -  b_\otimes^\dag  b_\otimes  \ , 
\label{Q}\\
U &=& b_\oplus^\dag  b_\otimes +  b_\otimes^\dag  b_\oplus  \ , 
\label{U}\\
V &=& i b_\oplus^\dag  b_\otimes - i  b_\otimes^\dag  b_\oplus  \ ,
\label{V}
\end{eqnarray}
where $Q,U$ represent the linear polarization and $V$ represents the circular polarization. 
The total number operator commutes with other operators such as
\begin{eqnarray}
[ N  , Q] = [ N  , U] =[ N  , V] =0 \ .
\end{eqnarray}
If we define $J_3 = Q/2, J_1 = V/2, J_3 =U/2$, we obtain
\begin{eqnarray}
[ J_i  , J_j] = i \epsilon_{ijk} J_k \ ,
\end{eqnarray}
where $\epsilon_{ijk}$ is the Levi-Civita symbol. We see an interesting connection between algebra of angular momentum and the algebra of two independent polarizations.
Thus, $Q,U,V$ consist of $SU(2)$ generators. 
It is useful to notice that these Stokes operators act on the operator $A_{\alpha,\beta}^+$ as
\begin{eqnarray}
[ N  , A_{\alpha ,\beta}^{\dag} ] = A_{\alpha ,\beta}^{\dag} \ ,\quad
[ Q  , A_{\alpha ,\beta}^{\dag} ] = A_{\alpha ,-\beta}^{\dag} \ ,\quad
[ U  , A_{\alpha ,\beta}^{\dag} ] = A_{\beta ,\alpha}^{\dag} \ ,\quad
[ V  , A_{\alpha ,\beta}^{\dag} ] = A_{i\beta , -i\alpha}^{\dag} \ ,
\label{tool}
\end{eqnarray}
Using these relations, we can calculate the expectation value of the intensity
\begin{eqnarray}
\langle\Psi , \alpha ,\beta| N |\Psi , \alpha ,\beta\rangle =
\langle\Psi , \alpha ,\beta| [ N  , \Psi  A_{\alpha ,\beta}^{\dag} ]  |\Psi , \alpha ,\beta\rangle = |\Psi|^2 
\end{eqnarray}
where we used Eqs.~(\ref{schrodinger}), (\ref{displacement}) and (\ref{tool}).
Similarly, the expectation values of the Stokes operators are as follows
\begin{eqnarray}
&&\langle\Psi , \alpha ,\beta| Q |\Psi , \alpha ,\beta\rangle
= |\Psi|^2\left(|\alpha|^2 -|\beta|^2 \right) \ , \\
&& \langle\Psi , \alpha ,\beta| U |\Psi , \alpha ,\beta\rangle 
= |\Psi|^2 \left(\alpha^* \beta  +  \beta^*\alpha\right)\ ,\\
&& \langle\Psi , \alpha ,\beta| V |\Psi , \alpha ,\beta\rangle 
= |\Psi|^2 \left( i\alpha^* \beta - i\beta^*\alpha  \right)\ ,
   \ .
\end{eqnarray}
where we used Eqs.~(\ref{creation}), (\ref{schrodinger}), (\ref{key}) and (\ref{tool}). The above expectation values describe the meaning of the coherent state $|\Psi , \alpha ,\beta\rangle$.
For example, if $\beta =0$, only $Q$ survives.
Hence, the state represents a linearly polarized graviton state. If $\beta =i \alpha $, only $V$ survives. Thus, the state represents a circularly polarized graviton state. In general, the state represents an elliptically polarized state.

It is straightforward to check an identity
\begin{eqnarray}
\langle\Psi , \alpha ,\beta| N |\Psi , \alpha ,\beta\rangle^2 =
\langle\Psi , \alpha ,\beta| Q |\Psi , \alpha ,\beta\rangle^2 
+ \langle\Psi , \alpha ,\beta| U |\Psi , \alpha ,\beta\rangle^2 
+ \langle\Psi , \alpha ,\beta| V |\Psi , \alpha ,\beta\rangle^2 
   \ .
\end{eqnarray}

Now let us discuss observations of primordial gravitational waves in the polarized initial state $|\Psi , \alpha ,\beta\rangle$. In this paper, we assume that the initial coherent state is generated by a consequence of some dynamics before inflation and we will not study the concrete mechanism for it. Once the concrete mechanism is specified, we would be able to specify the parameters $\alpha$, $\beta$ and $\Psi$. The dynamical generation of polarization of primordial gravitational waves has been discussed in the literature~\cite{Takahashi:2009wc,Satoh:2007gn,Satoh:2008ck,Obata:2016tmo,Obata:2016xcr,Soda:2011am}. The point to generate polarizations is to break the de Sitter invariance of the background spacetime. 

If we use Eq.~(\ref{operatorsI}), the power spectrum of primordial gravitational waves for the plus mode is evaluated as
\begin{eqnarray}
\langle\Psi , \alpha ,\beta| h_\oplus^{\bm k} h_\oplus^{\bm k}|\Psi , \alpha ,\beta\rangle 
&=&  \langle\Psi , \alpha ,\beta|\left\{ 
|v_k(\eta)|^2 + 2b^{{\bm k}\dag}_{\oplus} b_\oplus^{\bm k} |v_k(\eta)|^2
+ v_k(\eta)^2 b_\oplus^{\bm k} b_\oplus^{-\bm k} 
+ v^*_k(\eta)^2 b^{{\bm k}\dag}_{\oplus}b^{-{\bm k}\dag}_{\oplus}
\right\}|\Psi , \alpha ,\beta\rangle  \nonumber\\
&=& |v_k(\eta)|^2+ 2 |\Psi|^2 |\alpha|^2  |v_k(\eta)|^2
     + \Psi^2 \alpha^{*2}  v_k(\eta)^2 
     +\Psi^{*2} \alpha^2  v^*_k(\eta)^2
   \ ,
\end{eqnarray}
where we used the relations in (\ref{key}).
Similarly, we obtain the power spectrum for the cross mode
\begin{eqnarray}
\langle\Psi , \alpha ,\beta| h_\otimes^{\bm k} h_\otimes^{\bm k}|\Psi , \alpha ,\beta\rangle
&=&  \langle\Psi , \alpha ,\beta|\left\{ 
|v_k(\eta)|^2 + 2b^{{\bm k}\dag}_{\otimes} b_\otimes^{\bm k} |v_k(\eta)|^2
+ v_k(\eta)^2 b_\otimes^{\bm k} b_\otimes^{-\bm k} 
+ v^*_k(\eta)^2 b^{{\bm k}\dag}_{\otimes}b^{-{\bm k}\dag}_{\otimes}
\right\}|\Psi , \alpha ,\beta\rangle  \nonumber\\
 &=& |v_k(\eta)|^2+ 2 |\Psi|^2 |\beta|^2  |v_k(\eta)|^2
     + \Psi^2 \beta^{*2}  v_k(\eta)^2 
     +\Psi^{*2} \beta^2  v^*_k(\eta)^2 
\end{eqnarray}
and the power spectrum between the plus and cross modes
\begin{eqnarray}
\langle\Psi , \alpha ,\beta| h_\oplus^{\bm k} h_\otimes^{\bm k}|\Psi , \alpha ,\beta\rangle 
&=&  \langle\Psi , \alpha ,\beta|\left\{ 
 \left( b^{{\bm k}\dag}_{\otimes} b_\oplus^{\bm k} 
 +b^{{\bm k}\dag}_{\oplus} b_\otimes^{\bm k} 
 \right)|v_k(\eta)|^2
+ v_k(\eta)^2 b_\otimes^{\bm k} b_\oplus^{-\bm k} 
+ v^*_k(\eta)^2 b^{{\bm k}\dag}_{\oplus}b^{-{\bm k}\dag}_{\otimes}
\right\}|\Psi , \alpha ,\beta\rangle  \nonumber\\
&=& |\Psi|^2  |v_k(\eta)|^2 
\left( \alpha^* \beta +  \beta^* \alpha \right)
     + \Psi^2  v_k(\eta)^2 \alpha \beta
     +\Psi^{*2}   v^*_k(\eta)^2  \alpha^*\beta^*
   \ .
\end{eqnarray}
In principle, we can observe these power spectrum by using detectors of future  gravitational wave interferometer.
In particular, it would be a gravitational wave detector for high frequencies to detect the 
gravitons~\cite{Ito:2019wcb}. In our analysis, we assumed that the mode function $v_k$ is polarization independent 
but the polarization dependence of $v_k^{\rm A}$ will be necessary to make the final predictions.

\section{Conclusion}

In conventional inflationary scenario, we choose the Bunch-Davies vacuum as the initial state and discuss the polarizations of primordial gravitational waves in modified gravity or by using matter fields that survive during inflation. In this paper, however, we  considered Schr\"odinger type coherent state constructed from $SU(2)$ coherent states as the initial state~\cite{Moran, Hussin}. Since the state incorporates the degrees of polarizations, we used it as the polarized initial state of primordial gravitational waves without considering modified gravity or matter fields during inflation.  We then calculated the power spectrum of the  primordial gravitational waves for observations.
 
In this paper, we simply assumed the polarized state as the initial state and did not specify any mechanism to generate the polarized initial state. 
However, once the concrete mechanism is specified, we would be able to specify the parameters $\alpha$, $\beta$ and $\Psi$ in the polarized initial state $|\Psi ,\alpha,\beta\rangle $. In the history of the universe, the initial polarized state becomes a squeezed coherent state in the radiation dominated era. Then the graviton statistics tells us that primordial gravitational waves have a chance to keep their nonclassicallity~\cite{Kanno:2018cuk, Kanno:2019gqw}. Furthermore, it is known that  Hanbury-Brown-Twiss interferometry makes it possible to distinguish the nonclassicality~\cite{HanburyBrown:1956bqd,Brown:1956zza}. Related to this topic, the pioneering attempts were made to understand the quantum coherence~\cite{Giovannini:2010xg,Giovannini:2017uty,Giovannini:2017uty}. In the context of cosmic microwave radiations, the application of the Hunbury-Brown-Twiss interferometery was considered in~\cite{Chen:2017cgw}. Since the squeezed state is an entangled state, the entanglement between the modes of polarizations could be generated. Then we would be able to characterize the polarized initial state by calculating various entanglement measures as in~\cite{Maldacena:2012xp,Kanno:2014lma,Iizuka:2014rua,Kanno:2017dci,Kanno:2016gas,Kanno:2014bma}. We leave these issues for future works.

\begin{acknowledgments}
S.~K. was supported by JSPS KAKENHI Grant No. JP19K03827. J.~S. was in part supported by JSPS KAKENHI
Grant Numbers JP17H02894, JP17K18778, JP15H05895, JP17H06359, JP18H04589.
J.~S. was also supported by JSPS Bilateral Joint Research
Projects (JSPS-NRF collaboration) `` String Axion Cosmology.''
This research was supported by the Munich Institute for Astro- and Particle Physics (MIAPP) which is funded by the Deutsche Forschungsgemeinschaft (DFG, German Research Foundation) under Germany's Excellence Strategy – EXC-2094 – 390783311.
\end{acknowledgments}

\end{document}